\documentclass[letterpaper,fleqn]{cas-dc}
\usepackage[numbers,sort]{natbib}

\usepackage{flushend}

\begin{document}
\let\WriteBookmarks\relax
\def\floatpagepagefraction{1}
\def\textpagefraction{.001}

\title[mode = title]{{\normalsize Letter to the Editor}\\[5mm] \Large Comment on: ``Stokes' first problem for heated flat plate with Atangana--Baleanu fractional derivative'' [Chaos Solitons Fractals 117 (2018) 68]}
\shorttitle{Comment on: ``Stokes' first problem for heated flat plate...''}

\author[1]{Ivan~C.\ Christov}[orcid=0000-0001-8531-0531]
\ead{christov@purdue.edu}
\ead[url]{tmnt-lab.org}
\shortauthors{I.~C.~Christov}

\address[1]{School of Mechanical Engineering, Purdue University, West Lafayette, Indiana 47907, USA}

\begin{abstract}
In the sense of distributions, the derivative of the Heaviside unit step function $H(t)$ is a generalized Dirac-$\delta$ distribution. If the velocity $V(t)$ of a flat plate is impulsive, as $V(t)=H(t)$ (i.e., it is suddenly set into motion with unit velocity at $t=0^+$), then its acceleration is $V'(t)=\delta(t)$. The Dirac-$\delta$ distribution has no point values. However, when the Dirac-$\delta$ is the forcing term of an ODE (in $t$), it contributes to the solution. The recently published paper [Chaos Solitons Fractals 117 (2018) 68] incorrectly treats the Dirac-$\delta$ function as being identically 0. This Comment analyzes the source of this error, and provides guidance on how to correct it (based on the established  literature). The mathematical error identified is in addition to some issues about rheological models with fractional derivatives, which are also noted. That is to say, whether or not an ``Atangana--Baleanu fractional derivative'' is used in [Chaos Solitons Fractals 117 (2018) 68], the solution to Stokes' first problem provided therein is not correct.
\end{abstract}


\maketitle


Previously, there have been a number of unsuccessful attempt to solve Stokes' first problem of the impulsively moved plate for certain viscoelastic fluids \cite{C10}. The mathematical challenge arises in the fact that these certain viscoelastic fluids require the  time-differentiation of the impulsive boundary condition, which states that the dimensionless fluid velocity $u$ obeys $u(y=0,t)=H(t)$ at the plate $y=0$, where $H(t)$ is Heaviside's unit step function. Indeed, it is well known from the theory of generalized functions \cite[p.~39]{Kanwal} that $dH/dt = \delta(t)$, i.e., the Dirac-$\delta$ distribution (see, e.g., \cite{CJ09} and the references therein). In the sense of distributions, $\delta(t)$ does not possess point values and, most certainly does not $=0\;\forall t$ (see, e.g., \cite{C11} and the references therein).

Despite this issue being addressed in significant  detail in the literature  \cite{J05,CJ09,CC10,C10,J10_NA}, a recently published paper \cite{Sene19} in \textit{Chaos, Solitons \& Fractals} also commits this mathematical error.

Specifically, following the recommendation of the conclusions in \cite{CJ12}, we specialize the Fourier--Laplace (dual transform, $(y,t)\mapsto(w,s)$) domain solution from \cite[Eq.~(33)]{Sene19} to the second-grade fluid (take  $\alpha=1\Rightarrow\lambda=0$ in \cite{Sene19}):
\begin{equation}
\label{wrong}
    \bar{u}_s(w,s) = \sqrt{\frac{2}{\pi}} w \left\{\frac{1}{s[s + w^2\lambda_r s + w^2]} \right\}
\end{equation}
Compare the last equation to \cite[Eq.~(5)]{C11} (taking the relaxation time therein to zero, adjusting notation and non-dimen\-sionalizing):
\begin{equation}
\label{right}
    \bar{u}_s(w,s) = w\sqrt{\frac{2}{\pi}} \left\{\frac{1 + {\color{red}\lambda_r s}}{(1 + \lambda_r w^2)s + w^2}\right\}\frac{1}{s},
\end{equation}
which is the established, correct dual transform solution. 

Comparing Eqs.~\eqref{right} and \eqref{wrong}, we observe that \cite[Eq.~(33)]{Sene19} is missing the term `$\lambda_r s$'. This error arises when going from the first equality in the unnumbered equation above~ \cite[Eq.~(28)]{Sene19}, in which $u(0,t)=H(t)$ appears, to the second equality, in which $H(t)$ is `lost.' This is a nontrivial issue because, next the expression is substituted into \cite[Eq.~(28)]{Sene19}, wherein a time derivative of $H(t)$ would have to be taken (the last term). The time derivative of $H(t)$ (whether integer or `fractional') is not the same as the time derivative of `$1$' (whether integer or `fractional').

Thus, \cite[Eq.~(33)]{Sene19} (on which the remainder of \cite{Sene19} is based) is wrong. An independent way to establish the same conclusion (that \cite[Eq.~(33)]{Sene19}, on which the remainder of \cite{Sene19} is based, is wrong) is to consider the comment in \cite[Sec.~5]{Sene19}. It is stated that the time-domain solution \cite[Eq.~(42)]{Sene19} agrees with a prior solution from the literature. Actually said solution, which is not from Ref.~[30] in \cite{Sene19}, is erroneous and was corrected in \cite{CC10}. It follows that, since the solution for Stokes' first problem in \cite{Sene19} is incorrect for $\alpha=1$, then it is also incorrect $\forall\alpha\in[0,1]$. (That is to say, it is sufficient to provide one counterexample to disprove the general statement that a correct solution was obtained in \cite{Sene19} $\forall$ fractional powers $\alpha\in[0,1]$.)  

Additionally, since the thermal problem's solution in \cite[Sec.~8]{Sene19} relies on the incorrect flow solution from earlier sections, then the calculations in \cite[Sec.~8]{Sene19} will have to be revisited and corrected as well.

It is also worth pointing out that the figures provided in \cite{Sene19} do not show any of the effects of posing a model with a fractional derivative because Figs.~1--3 are for $\alpha=0$ (Newtonian fluid) and Figs.~4--6 are for $\alpha=1$ (classical second-grade fluid with integer-order derivative in the constitutive law)

On the basis of this discussion, we also observe that Refs.\ [11,12,14,19,30,38] in \cite{Sene19} are papers that have been proven to be mathematically incorrect, as discussed in \cite{C10,JP04,C11}. Therefore, the reader is cautioned not to refer to these works  cited in \cite{Sene19} without consulting the latest developments in the literature \cite{C10,JP04,C11}.

Finally, a note of caution is warranted regarding the proposed rheological fluid model in \cite[Eqs.~(11--13)]{Sene19}. While the so-called second-grade fluid is derived on the basis of general notions of continuum mechanics \cite{ER54}, replacing the time derivative therein with a ``fractional derivative'' (a temporal integro-differential operator with certain properties) is not justified per the development of the Rivlin--Ericksen tensors and the order-$n$ expansion. Perhaps the approximation theorem for functionals due to Coleman and Noll \cite{Cono} can be revisited to yield some kind of hereditary integral like \cite[Eq.~(4)]{Sene19}. However, at the time of this submission, there is no evidence known to the author that such a calculation is possible. Thus, we must conclude that the derivation of the second-grade fluid's rheological model does not allow for a replacement of integer-order derivatives by `fractional' ones simply on the \emph{basis of analogy}. It follows that the real-world relevance of the incorrectly-solved model proposed in \cite{Sene19} has not been established. An experimental verification of the ``generalized second-grade fluid with a new fractional derivative operator'' should first be conducted to determine the model's validity.

\bibliographystyle{cas-model2-names}

\bibliography{csf_refs.bib}

\end{document}